\documentclass[10pt]{article}
\usepackage[utf8]{inputenc}
\usepackage{graphicx}
\usepackage{amsmath}
\usepackage{hyperref}
\usepackage{geometry}
\title{Geospatial and Symbolic Hypothesis for the Foundation of Tenochtitlan Based on Digital Elevation Analysis of the Valley of Mexico}
\author{
  José Alberto Baeza Guerra \\
  \texttt{jose.alberto.baeza@gmail.com} \\
  \url{https://www.linkedin.com/in/josealbertobaeza/} \\
  GitHub: \url{https://github.com/jsonata}
}
\date{April 2025}

\begin{document}

\maketitle

\begin{abstract}
This paper proposes a novel hypothesis about the foundation of Tenochtitlan by combining digital elevation modeling with historical and symbolic analysis. Using geospatial data from EarthExplorer, we simulate various historical water levels in the Valley of Mexico. The resulting lake configurations reveal possible locations for ancient settlements near now-vanished shorelines, suggesting a dynamic transformation of sacred geography that aligns with key Mexica myths. We identify Santa María Aztahuacan as a strong candidate for the historical Aztlan and propose a reinterpretation of foundational codices in light of geomythical correlations.
\end{abstract}

\section{Introduction}
Most modern reconstructions of the Valley of Mexico's lakes are based on the state of the basin at the time of the Spanish conquest or shortly thereafter. However, these reconstructions might not reflect the hydrological reality of earlier centuries. In this study, we explore higher possible lake levels using satellite elevation data to reassess ancient geography and propose symbolic interpretations rooted in Mexica cosmology.

Aztlan, traditionally considered the mythical origin place of the Mexica people, remains one of the most enigmatic topics in Mesoamerican studies. Despite its prominent role in origin myths and codices, the actual geographic location of Aztlan has never been conclusively identified. Over the years, numerous theories have been proposed, ranging from sites in the north of Mexico (such as Nayarit, Aztatlán) to locations closer to the Valley of Mexico, or even symbolic non-places representing spiritual origins \cite{smith1984aztlan, munoz2019aztlan, thoughtcoAztlan}.

To the best of our knowledge, there are no prior academic proposals identifying Santa María Aztahuacan as a candidate for the mythical Aztlan. However, the site's etymology ("place of herons"), its emergence as an island in paleo-lake reconstructions, and its archaeological antiquity suggest it deserves consideration within the broader discourse surrounding the origin of the Mexica migration.

The historical ambiguity surrounding Aztlan underscores the importance of interdisciplinary approaches that combine geography, symbolism, and cultural memory. This work contributes to that ongoing search by suggesting a possible identification of Aztlan based on geospatial modeling and topographic transformation of the valley's lake system.

\section{Methodology}
Using TIF elevation files from EarthExplorer, we developed a JavaScript-based visualization tool that simulates lake coverage at varying altitudes. By adjusting the water level parameter, we identified patterns in shoreline emergence and submergence over time. The complete source code is publicly available at: \url{https://github.com/jsonata/mexica-geospatial-model}.

\section{Data and Tools}
Elevation data was obtained from the USGS EarthExplorer website (\url{https://earthexplorer.usgs.gov/}) by downloading multiple TIF tiles covering the Valley of Mexico. A custom HTML file was developed using JavaScript along with the Leaflet and GeoTIFF libraries. This setup enabled the dynamic rendering of a digital lake at different water levels over an OpenStreetMap base layer, using a canvas overlay. Specific markers for ancient and modern locations discussed in this work were added to the visualization.

To draw symbolic geodesic lines between points of interest, Google My Maps was employed. This allowed precise tracing of connections such as Tenayuca–Culhuacan and Tepeyac–Churubusco. Screenshots of the visualizations will be included to support the analysis.

\section{Results: Ancient Settlements and Shorelines}
At a water level of approximately 2257 meters above sea level, the lake covered a much larger area. Major ancient settlements such as Cuicuilco and Teotihuacan would have been located near the lake’s edge. Likewise, Tenayuca, Tepeyac, and Chapultepec—some of the oldest ceremonial centers—emerge as coastal sites. Tenochtitlan and Tlatelolco, by contrast, were not yet islands at this stage.

\begin{figure}[h!]
  \centering
  \includegraphics[width=0.9\textwidth]{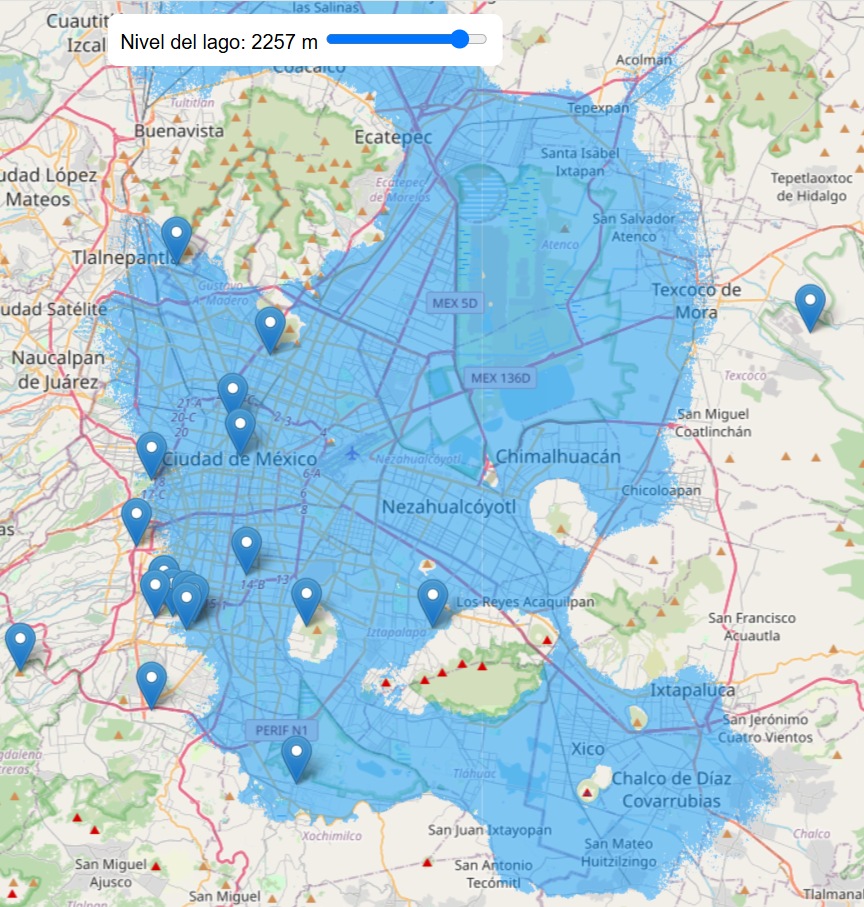}
  \caption{Simulation of the lake at 2257 meters above sea level. Ancient settlements appear near the shoreline.}
  \label{fig:lake2257}
\end{figure}

\begin{figure}[h!]
  \centering
  \includegraphics[width=0.9\textwidth]{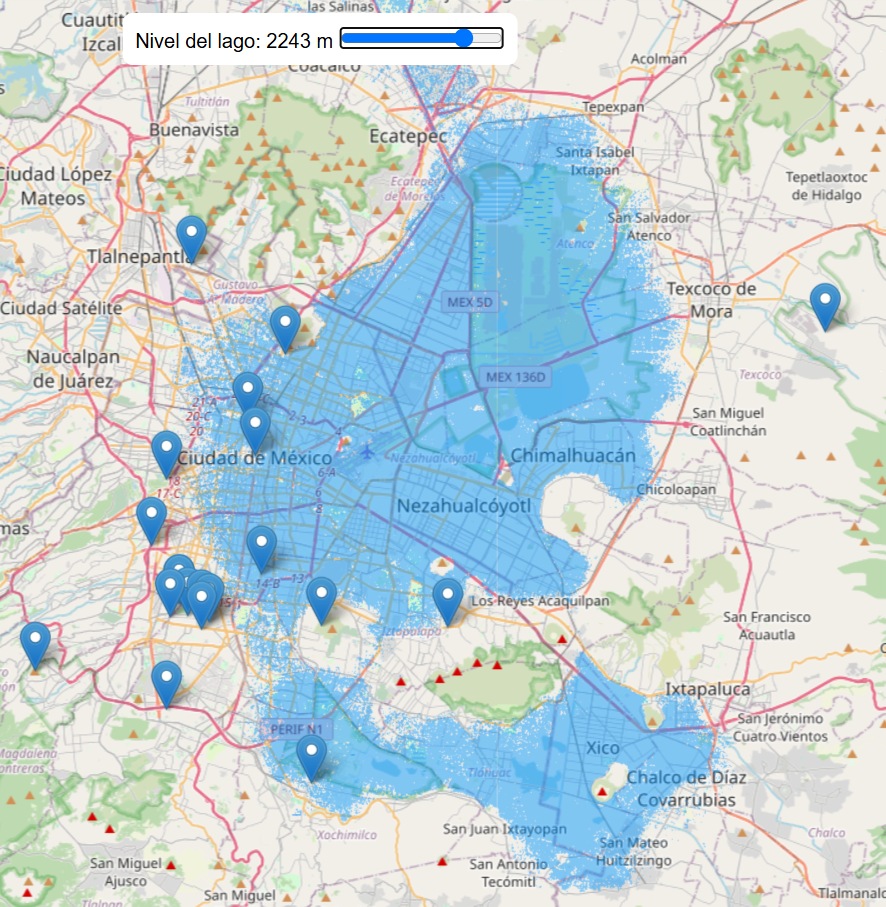}
  \caption{Lake at 2243 meters. New islands emerge, including Huitzilopochco.}
  \label{fig:lake2243}
\end{figure}

\begin{figure}[h!]
  \centering
  \includegraphics[width=0.9\textwidth]{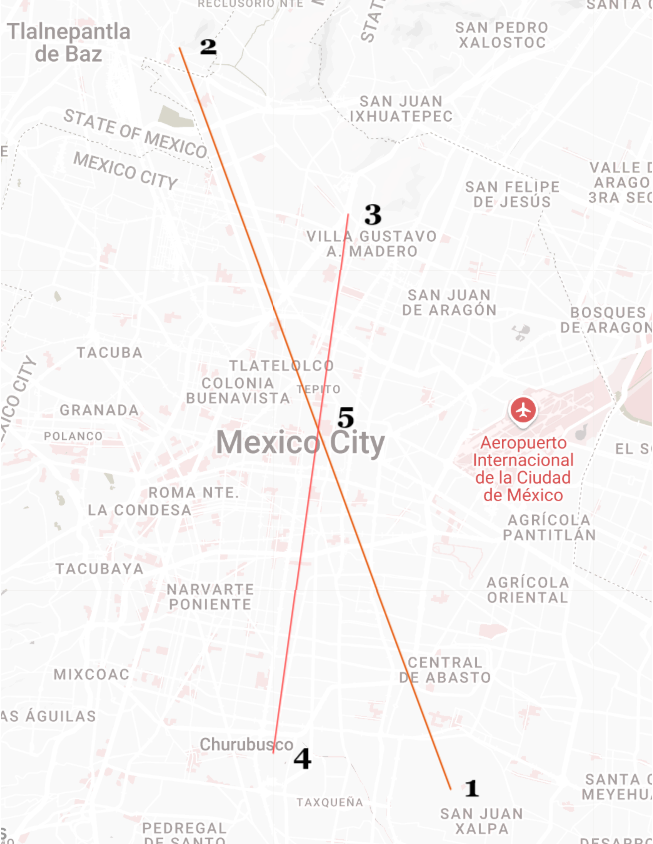}
  \caption{Symbolic geodetic lines intersecting at the Templo Mayor. Numbers indicate: (1) Cerro de la Estrella (Culhuacan), (2) Tenayuca, (3) Tepeyac, (4) island of Huitzilopochco, (5) intersection at the Templo Mayor.}
  \label{fig:crossing}
\end{figure}

\section{Aztahuacan as Aztlan}
Under the high-water model, the Cerro de la Estrella (Culhuacan) becomes an island. In front of it, a larger island aligns with present-day Santa María Aztahuacan, a name that shares the etymological meaning "place of herons" with Aztlan. This suggests a reinterpretation: that Santa María Aztahuacan could be the historical Aztlan.

Importantly, Santa María Aztahuacan is also one of the oldest continuously inhabited areas in the Mexican highlands. In 1953, human fossils were discovered there and dated to over 12,000 years ago, indicating a deep historical occupation of the site. This adds further weight to the hypothesis that Aztahuacan held long-term symbolic and practical significance for ancient populations of the region \cite{romano1953aztahuacan}.

This hypothesis is further supported by iconographic analysis. The "Tira de la Peregrinación" (Codex Boturini), which narrates the Mexica departure from Aztlan, depicts a group of people leaving an island and approaching a place that strongly resembles Culhuacan, where Huitzilopochtli—still unborn—is represented speaking from within a cave. Within the context of this study, this cave and its associated symbolism align with the high-water geography, where Culhuacan (Cerro de la Estrella) was indeed an island. The proximity of Santa María Aztahuacan across the water lends further plausibility to its identification as the mythical Aztlan, reinforcing both the geographical and narrative coherence of the proposed model \cite{tiraPeregrinacion}.

Moreover, in the first page of the "Tira de la Peregrinación", the year date "1 Flint" (1 Tecpatl) appears prominently, which matches the year of the Mexica departure from Aztlan. This chronological marker not only anchors the beginning of the migration but also resonates with other codices such as the Codex Mendoza, where it is absent from the founding cycle—suggesting a symbolic bridge between departure and foundation.

\section{Symbolic Geography and Myth}
\subsection{The Myth of Huitzilopochtli: A Brief Narration}
According to Mexica mythology, the goddess Coatlicue became miraculously pregnant after a ball of feathers fell into her lap. Her daughter Coyolxauhqui, along with her four hundred brothers (the Centzonhuitznahua), planned to kill her to prevent the birth. However, Coatlicue gave birth to Huitzilopochtli, the god of the sun and war, who emerged fully armed and killed Coyolxauhqui, dismembering her body, and casting his brothers into the sky as stars \cite{mitoHuitzilopochtli}.

This myth is interpreted as an allegory of the daily cosmic struggle: the triumph of the sun over the moon and stars. In this study, we reinterpret this myth as a symbolic reflection of geological and astronomical events in the Valley of Mexico.

During the desiccation process of the lake, several islands emerged. Among them was Huitzilopochco—later known as Churubusco—which translates to "place of Huitzilopochtli." Today, this area corresponds to the surroundings of Calzada de Churubusco, near the National Center for the Arts (CNA) and a golf course known as the Country Club. Its emergence during the lowering of water levels metaphorically reflects the birth of Huitzilopochtli from the sacred geography.

The name Huitzilopochtli itself means "left-handed hummingbird"—a solar metaphor. The hummingbird, with a disproportionately large heart and vibrant colors, was associated with the sun. The term "left-handed" reflects the sun’s path in the northern hemisphere, which tends to the southern sky—symbolically, the left.

Additionally, the Sierra de Guadalupe is interpreted in this framework as the embodiment of Coatlicue, the earth mother. This mountainous formation features two land extensions that entered into the ancient lake, marking the locations of Tenayuca and Tepeyac at their respective tips. Fertility rituals were reportedly conducted in this region, further reinforcing its association with maternal symbolism.

During the colonial period, Spanish ecclesiastical authorities constructed a church at Tepeyac, dedicated to the Virgin of Guadalupe—a name that bears striking phonetic and symbolic resemblance to Coatlicue. To this day, the site receives millions of pilgrims annually on December 12, a date which, prior to calendar reform, coincided with the winter solstice. For Mesoamerican peoples, the solstice marked the symbolic rebirth of the sun, aligning with themes central to the myth of Huitzilopochtli.

\section{Astronomical Marker: The 1325 Eclipse}
This geological transformation aligns with a solar eclipse observed in the Valley of Mexico on April 21, 1325, between 11:00 and 11:06 a.m., as documented by archaeoastronomer Jesús Galindo Trejo. The eclipse would have been visible from the area that would become Tenochtitlan and could have been interpreted by the Mexica as a divine omen. According to Galindo Trejo, after the eclipse, the Mexica may have waited two 13-day ritual periods (26 days total), culminating in the zenithal passage of the sun on May 17, 1325—an ideal symbolic moment to found the city. This celestial alignment supports the hypothesis that cosmological events played a guiding role in the Mexica's decision-making process \cite{galindoTrejo1325eclipse}.

In this work, we interpret the 1325 eclipse as the culmination of the Mexica mythical journey that began in Aztlan, where the sun had not yet been born and was symbolically represented as speaking from within a cave in Culhuacan. As the lake receded and new islands emerged, including Huitzilopochco (Churubusco), the sun metaphorically "rose" with the emergence of this sacred site. The total eclipse thus represents the final mythical battle between Huitzilopochtli (the sun) and Coyolxauhqui (the moon), in which the sun emerges victorious, signaling the appropriate time and place to establish Tenochtitlan.

\section{Interpreting the Codex Mendoza}
Research by archaeoastronomer Dr. Arturo Montero has identified a geodetic alignment linking the major temples of Tenayuca, Tlatelolco, and the Templo Mayor in Tenochtitlan. Extending this line southward reaches the Cerro de la Estrella (Huixachtécatl), reinforcing the notion of intentional spatial and symbolic planning \cite{monteroLagoTexcoco}.

Strikingly, these alignments correspond with present-day avenues in Mexico City, many of which were laid out along the original pre-Hispanic city structure. Additionally, we propose a second geodetic alignment, connecting the Tepeyac (symbolizing Coatlicue) with the former island of Huitzilopochco (symbolizing Huitzilopochtli). The fact that these two lines intersect precisely at the Templo Mayor strengthens the interpretation of the temple as the cosmological center of Mexica sacred geography.

In the Codex Mendoza \cite{codiceMendoza}, the founding date of Tenochtitlan is surrounded by 51 year symbols—one short of the 52-year cycle. The missing year, 1 Flint (1 Tecpatl), matches the departure year from Aztlan, indicating the closure of a cosmic cycle.

\section{Conclusion}
This interdisciplinary approach offers a novel interpretation of Mexica mytho-history grounded in hydrological, geographical, and astronomical data. The island emergence pattern aligns with the myth of Huitzilopochtli's birth, and the placement of ancient sites supports a reevaluation of Aztlan’s location. The intersection of symbolic lines reinforces the sacred centrality of Tenochtitlan. This research coincides with the 700th anniversary of Tenochtitlan’s founding, offering new insights into its origins.


\begin{thebibliography}{9}
\bibitem{smith1984aztlan} 
Smith, M. E., 
\textit{The Aztlan Migrations of the Nahuatl Chronicles: Myth or History?}, 
Ethnohistory, Vol. 31, No. 3 (1984), pp. 153–186. \url{https://www.jstor.org/stable/482619}

\bibitem{munoz2019aztlan} 
Muñoz-Hunt, T., 
\textit{Aztlán: From Mythos to Logos in the American Southwest}, 2019. \url{https://www.researchgate.net/publication/337752143_Aztlan_From_Mythos_to_Logos_in_the_American_Southwest}

\bibitem{thoughtcoAztlan} 
ThoughtCo, 
\textit{Aztlan, The Mythical Homeland of the Aztec-Mexica}. \url{https://www.thoughtco.com/aztlan-the-mythical-homeland-169913}

\bibitem{romano1953aztahuacan} 
Arturo Romano Pacheco, 
\textit{Nota preliminar sobre los restos humanos sub-fósiles de Santa María Aztahuacán, D. F.}, 
Anales del Instituto Nacional de Antropología e Historia, 1953. Available online: \url{https://www.revistas.inah.gob.mx/index.php/anales/article/view/7199/8042}

\bibitem{tiraPeregrinacion} 
Tira de la Peregrinación (Códice Boturini), INAH, Biblioteca Nacional de Antropología e Historia. Digital version available at: \url{https://www.mexicana.cultura.gob.mx/es/repositorio/detalle?id=_suri:INAH:TransObject:5cddfe477a8a022ef8480f30}

\bibitem{mitoHuitzilopochtli} 
\textit{El nacimiento de Huitzilopochtli}, 
Arqueología Mexicana. Available online: \url{https://arqueologiamexicana.mx/mexico-antiguo/el-nacimiento-de-huitzilopochtli}

\bibitem{galindoTrejo1325eclipse} 
Jesús Galindo Trejo, 
\textit{La luna cubrió el sol en 1325: posible señal de la fundación de México-Tenochtitlan}, 
Gaceta UNAM, 2022. Available online: \url{https://www.gaceta.unam.mx/la-luna-cubrio-el-sol-en-1325-posible-senal-de-la-fundacion-de-mexico-tenochtitlan/}

\bibitem{monteroLagoTexcoco} 
Arturo Montero García, 
\textit{El Lago de Texcoco y México-Tenochtitlan: 1519–1521}, 2019. Available online: \url{https://fliphtml5.com/oedap/rudg/El_Lago_de_Texcoco_y_M%C3%A9xico-Tenochtitlan%3A_1519-1521/}

\bibitem{codiceMendoza} 
Códice Mendoza, ca. 1541. Bodleian Library, Oxford. Digital facsimile available at: \url{https://digital.bodleian.ox.ac.uk/objects/2dc6f6e9-0e12-446e-9d2e-0ba72d4f9a2e/}
\end{thebibliography}
\end{document}